\documentclass[aps,pra,twocolumn]{revtex4-1}
\usepackage{amsmath,amsfonts,amssymb}
\usepackage[caption=false]{subfig}
\usepackage{pict2e}

\usepackage[usenames,dvipsnames]{color}

\newcommand{\ket}[1]{\ensuremath{|#1\rangle}}
\newcommand{\bra}[1]{\ensuremath{\langle#1|}}
\newcommand{\ketbra}[2]{\ensuremath{\ket{#1}\!\bra{#2}}}

\newcommand{\In}{\ensuremath{\text{in}}}
\newcommand{\Out}{\ensuremath{\text{out}}}
\newcommand{\Hc}{\ensuremath{\text{H.c.}}}

\newcommand{\G}{\ensuremath{\mathcal{G}}}
\newcommand{\Z}{\ensuremath{\mathbb{Z}}}

\newcommand{\vin}{\ket{v_{\In}}}
\newcommand{\vout}{\ket{v_{\Out}}}

\newcommand{\vl}{\ensuremath{v_{\ell}}}
\renewcommand{\vr}{\ensuremath{v_{r}}}
\newcommand{\vt}{\ensuremath{v_{t}}}
\newcommand{\vb}{\ensuremath{v_{b}}}

\newcommand{\beq}[1]{\begin{equation}\label{#1}}
\newcommand{\beqs}{\begin{equation*}}
\newcommand{\eeq}{\end{equation}}
\newcommand{\eeqs}{\end{equation*}}

\newcommand{\e}{\ensuremath{\mathrm{e}}}

\renewcommand{\thicklines}{\linethickness{0.125em}}

\newcommand{\dashedLines}{\thicklines\color{red}}
\newcommand{\dottedLines}{\thicklines\color{blue}}
\newcommand{\solidLines}{\thicklines\color{ForestGreen}}

\newcommand{\solid}{\ensuremath{g}}
\newcommand{\dashed}{\ensuremath{r}}
\newcommand{\dotted}{\ensuremath{b}}
\newcommand{\Gsolid}{\ensuremath{\G_{\solid}}}
\newcommand{\Gdashed}{\ensuremath{\G_{\dashed}}}
\newcommand{\Gdotted}{\ensuremath{\G_{\dotted}}}
\newcommand{\GdashORdot}[1]{\ensuremath{\G_{#1}}}

\newcommand{\seq}{\ensuremath{\mathcal{S}}}

\begin{document}

\title{Universal quantum computation by discontinuous quantum walk}
\author{Michael S.\ Underwood}
\email{munderwood@qis.ucalgary.ca}
\author{David L.\ Feder}
\email[Corresponding author: ]{dfeder@ucalgary.ca}
\affiliation{Institute for Quantum Information Science,
University of Calgary, Alberta T2N 1N4, Canada}

\date{\today}

\begin{abstract}
Quantum walks are the quantum-mechanical analog of random walks, in which a 
quantum `walker' evolves between initial and final states by traversing the 
edges of a graph, either in discrete steps from node to node or via continuous 
evolution under the Hamiltonian furnished by the adjacency matrix of the graph.
We present a hybrid scheme for universal quantum computation in which a 
quantum walker takes discrete steps of continuous evolution. This 
`discontinuous' quantum walk employs perfect quantum state transfer between
two nodes of specific subgraphs chosen to implement a universal gate set,
thereby ensuring unitary evolution without requiring the introduction of an 
ancillary coin space. The run time is linear in the number of simulated qubits 
and gates. The scheme allows multiple runs of the algorithm to be executed
almost simultaneously by starting walkers one timestep apart.
\end{abstract}

\maketitle

\section{Introduction}
	
In analogy to the use of random walks to speed up classical
computation~\cite{Woeginger2003}, the role of quantum walks has been explored
in the realm of quantum
computation~\cite{Kempe2003-CP44-307,Ambainis2003-IJoQI1-507}.
Quantum walks were first applied to quantum algorithms known to be more
efficient than their classical counterparts, such as Grover's search of an
unsorted array~\cite{Shenvi2003-PRA67-52307,Santha2008}, the element
distinctness problem~\cite{Buhrman2001}, and triangle
finding~\cite{Ambainis2007-SJoC37-210}
and its extension to $k$-cliques~\cite{Childs2005-QIaC5-593}. It was quickly 
recognized that quantum walks could be also used to generate quantum algorithms
for various problems more directly than was possible within the context of the
conventional quantum circuit model, for example traversing glued binary 
trees~\cite{Childs2002-QIP1-35,Childs2003}, and evaluating decision
trees~\cite{Farhi1998-PRA58-915},
NAND trees~\cite{Farhi2008-ToC4-169,Childs2009-ToC5-119}
and game trees (AND-OR fomulas)~\cite{Reichardt-arxiv-0907.1623}. 

More recently, quantum walks have been shown to be computationally universal 
in both the continuous-time~\cite{Childs2009-PRL102-180501} and 
discrete-time~\cite{Lovett2010-PRA81-42330} formulations. In both cases, 
the walker moves from left to right along `rails' or lines of vertices, 
labelled by computational basis states. 
These rails are interspersed with small graphs, or `widgets,' that transform the
state of the quantum walker in analogy to gates in the circuit model. The 
widgets are attached either to individual rails or between pairs of rails, and 
the transformations are chosen in such a way as to effect a desired 
computation.  The collection of rails and widgets forms a computational graph,
which mimics the circuit model via the unitary evolution of the walker in its 
spatial Hilbert space.  

The continuous-time model for universal computation proposed in 
Ref.~\onlinecite{Childs2009-PRL102-180501} makes use of a walker with a tightly 
peaked momentum profile. This requires each of its rails to include semi-infinite
`tails' (linear graphs) both before and after the computational graph,
though in practice the length of these tails needs only to be large compared to 
twice the total evolution time of the walker within the graph (i.e.\ 
proportional to the circuit depth). 
Additionally, the preparation of the momentum state requires an initial 
sequence of momentum filter widgets, each with its own tail.
A side effect is that most of the walker's probability never enters the
computational graph. While these considerations only increase the resources
polynomially in the number of widgets used, their presence makes the scheme
somewhat cumbersome.

In the discrete-time scheme of Ref.~\onlinecite{Lovett2010-PRA81-42330}, 
double-edged rails are employed in order to guarantee that the walker moves
through the entire graph strictly from left to right. Each vertex is attached 
to four edges, two of which are connected to the vertex to the left, and two 
to the right. While this scheme does not require tails, the walker must have 
at least eight internal states because the various widgets require two-, four-,
and eight-dimensional coins. 

An alternative approach to universal quantum computation discussed in this work
is based on `perfect state transfer'
(PST)~\cite{Bose-ContempPhys48-13,Kay-arxiv-0903.4274v3}. In 
PST, quantum states are transferred perfectly between two nodes of a graph in 
continuous time. While PST was originally described in terms of spin 
chains~\cite{Christandl2004-PRL92-187902}, it has more recently been
extended to continuous-time quantum walks on 
graphs~\cite{Feder2006-PRL97-180502,Bernasconi-PRA78-052320,Severini-arXiv-1001.0674}.
In the latter formulation of PST, the walker's state at the output vertex is 
identical to that at the input, modulo a phase. In order to simulate a 
universal quantum circuit for $n$ qubits, one would need to construct a graph 
in which an input state on $2^n$ vertices could be transferred via PST to 
$2^n$ vertices, together with the desired $2^n$-dimensional unitary operator 
$U$. While this task appears to be difficult for general $U$, it might be
possible to decompose the graph into a small set of widgets, each of which 
individually allows for PST or a straightforward extension of it. Alternating
these widgets with some other set of processes could then result in universal 
quantum computation~\cite{Godsil-PRA81-052316-2010} in a manner analogous to
the interleaving of driving and query Hamiltonians that can efficiently
simulate continuous-time quantum query algorithms under the discrete query
model \cite{Cleve-STOC}.

We present a hybrid scheme for universal quantum computation that combines the 
best features of the continuous and discrete-time schemes discussed above 
while minimizing their disadvantages: the walker undergoes PST under 
continuous evolution, but only in discrete steps. In this `discontinuous
quantum walk,' a set of widgets are turned on and off at discrete time 
intervals in a prescribed manner. The walker moves through these graphs in 
sequence, resulting in the implementation of the desired $2^n$-dimensional 
unitary $U$. In the absence of errors, the initial state propagates through 
the graph without loss of amplitude at the output; furthermore, no coin degree 
of freedom is required even though the procedure utilizes discrete timesteps.
The scheme has the added advantage that new walkers can be sent through the 
same graph at regular intervals, allowing for nearly simultaneous repetition of
the algorithm with no additional overhead.

The remainder of this paper is organized as follows.  In 
Section~\ref{sec:mainScheme} we define the hybrid scheme for universal quantum 
computation via discontinuous quantum walk. In Section~\ref{sec:blocks} we 
describe a set of fundamental elements that fulfill the requirements of the 
scheme, and in Section~\ref{sec:universal} show how to combine them to create 
a universal set of gates. In Section~\ref{sec:summary} we provide some 
concluding remarks.

\section{Hybridizing Discrete-Time and Continuous-Time Walks\label{sec:mainScheme}}
A quantum walk takes place on a graph $\G=(V,E)$, where $V$ is a set of 
vertices and $E\subseteq V\times V\times W$ is a set of edges defined by pairs 
of elements of $V$ and associated edge weights $w_{ij}$ taken from 
$W=\{w_{ij}\}$. Often $W$ is simply the single-element set $\{1\}$, in which 
case it need not be present, but more generally it can be any set of numbers. 
An undirected weighted graph $\G$ is defined by a corresponding adjacency matrix
$G$, with matrix elements defined by
\beq{eq:adjacencyMatrix}
	G_{ij} = \left\{
		\begin{array}{cc}
			w_{ij}, & (i,j,w_{ij})\in E, \\
			0, & \text{otherwise},
		\end{array}\right.
\eeq
where $w_{ij}=w_{ji}>0$.
By definition, $G$ is real and symmetric, and can therefore be interpreted as 
a Hamiltonian on the state space $\mathcal{V}=\{\ket{v}:v\in V\}$.  Doing so 
describes a continuous-time quantum walk on \G, where a quantum walker initially
on  vertex $v_I$ in the state $\ket{I}=\ket{v_I}$ evolves in time $t$ to 
the final state $\ket{F}=\exp(-i G t)\ket{v_I}$ which is generally a 
superposition of vertex states $\ket{v}\in \mathcal{V}$. 

\begin{figure}
	\setlength{\unitlength}{0.042\textwidth}
	\subfloat{
		\label{fig:PSTline}
		\begin{picture}(11,0.9)(-0.8,-0.1)
			\put(-0.7,0.025){\footnotesize(a)}
			\multiput(0,0.1)(2,0){6}{\circle*{0.15}}
			\multiput(0,0.1)(6,0){2}{\line(1,0){4}}
			\put(0.4,0.3){$\scriptstyle\sqrt{1\cdot M}$}
			\put(2.15,0.3){$\scriptstyle\sqrt{2(M-1)}$}
			\put(6.15,0.3){$\scriptstyle\sqrt{(M-1)2}$}
			\put(8.4,0.3){$\scriptstyle\sqrt{M\cdot 1}$}
			\put(4.75,0){$\cdots$}
		\end{picture}
	}\\
	\subfloat{
		\label{fig:PSTdiscrete}
		\begin{picture}(11,0.6)(-0.8,-0.1)
			\put(-0.7,0.075){\footnotesize(b)}
			\multiput(0,0.2)(6,0){2}{\multiput(0,0)(1,0){5}{\circle*{0.15}}}
			\multiput(0,0.2)(6,0){2}{\multiput(0,0)(2,0){2}{\line(1,0){1}}}
			\multiput(1,0.2)(6,0){2}{
				\multiput(0,0)(2,0){2}{
					\multiput(0,0)(0.2,0){5}{\line(1,0){0.1}}
				}
			}
			\put(4.75,0.1){$\cdots$}
		\end{picture}
	}
	\caption{Two different methods to effect perfect state transfer.
			\subref{fig:PSTline} A line of $M$ segments weighted so as to provide
			perfect state transfer from one end to the other in time $\pi/2$.
			All line segments are always `on'
			but have varying weights $w_{j,j+1}=\sqrt{j(M+1-j)}$,
			$j\in\{1,\ldots,M\}$.
			\subref{fig:PSTdiscrete} A line of $M$ segments with unit weight, with
			the solid and dashed couplings turned on alternately so as to provide
			perfect state transfer from one end to the other in time $M\pi/2$.
			That is, the coupling between a pair of adjacent nodes is
			alternated between 0 and 1.}
\end{figure}
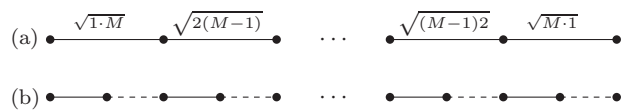

With perfect quantum state transfer (PST), the final state after a time 
$t_0>0$ corresponds to unit probability on a single vertex $v_F$, 
$\ket{F}=\ket{v_F}$. In particular, a line of $M$ segments exhibits PST 
from one end to the other for the particular choice of edge weights 
$w_{j,j+1}=\sqrt{j(M+1-j)}$
for $j\in\{1,\ldots,M\}$~\cite{Christandl2004-PRL92-187902};
an example is shown in 
Fig.~\ref{fig:PSTline}. In this situation, a walker initially localized at 
time $t=0$ on the left-most node, $\ket{I}=\ket{v_0}$, will be localized on 
the right-most node at time $t=\pi/2$; a phase of $(-i)^M$ will have been
applied to the state.
Consider instead 
the combination of line segments in Fig.~\ref{fig:PSTdiscrete}.  If the walker 
begins on the left-most node and the dashed line segments are disabled, i.e.\ 
their weight or coupling constant vanishes, then after a time $t=\pi/2$ the 
walker will have transferred perfectly to the second node and acquired a phase
of $-i$. If at this point the solid lines 
are switched off and the dashed ones enabled, the walker will proceed to the 
third node. In this manner it can be perfectly transferred to the right-most 
node in $M$ discrete steps, taking a total time of $M\pi/2$. Note that the 
scheme requires only two different unconnected graphs, those with solid and 
dashed edges shown in Fig.~\ref{fig:PSTdiscrete}, that are enabled in an 
alternating pattern. The direction the walker travels on these `transport'
rail segments then depends crucially on the initial occupied node.

The representation of a qubit requires two such rails, with one encoding the 
logical $\ket{0}$ state and the other the logical $\ket{1}$. Operations on the
qubit are effected by interspersing the transport segments with widgets
that transform the walker in non-trivial ways. To affect the 
relative phases of $\ket{0}$ and $\ket{1}$, equivalent to a rotation 
$R_Z(\theta)$ of the encoded qubit by an angle $\theta$ about the $Z$ axis, one 
needs to add an identity widget to the first rail and a phase widget to the
second. Both widgets must take the same amount of time to traverse by a 
continuous-time quantum walk, and both must have PST. For a universal
single-qubit
gate one also requires a rotation about an orthogonal axis $X$ or $Y$. This
requires a widget that connects the two rails, in such a way that after 
continuous evolution for a specified time the amplitude on the two rails will
have been transferred into a different superposition of $\ket{0}$ and
$\ket{1}$. 
This generalizes the concept of PST: arbitrary probability amplitude should 
remain on the input and output vertices, but no amplitude can remain on any
other vertex of the widget.

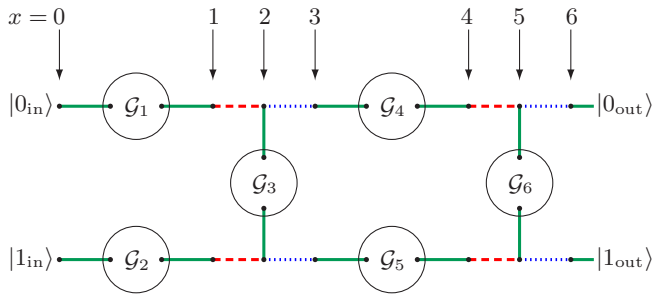
\begin{figure}
\setlength{\unitlength}{0.68cm}
		\begin{picture}(12.6,5.75)(-0.5,0.25)
			\put(-0.5,3.9){$\ket{0_{\In}}$}
			\put(-0.5,0.9){$\ket{1_{\In}}$}
			\put(11,3.9){$\ket{0_{\Out}}$}
			\put(11,0.9){$\ket{1_{\Out}}$}
			
			{\solidLines
			\multiput(0.5,1)(0,3){2}{\line(1,0){1}}
			\multiput(2.5,1)(0,3){2}{\line(1,0){1}}
			\multiput(5.5,1)(0,3){2}{\line(1,0){1}}
			\multiput(7.5,1)(0,3){2}{\line(1,0){1}}
			\multiput(10.5,1)(0,3){2}{\line(1,0){0.45}}
			\multiput(4.5,1)(0,2){2}{\line(0,1){1}}
			\multiput(9.5,1)(0,2){2}{\line(0,1){1}}
			}%
			
			{\dashedLines
			\multiput(3.5,1)(0.25,0){4}{\line(1,0){0.15}}
			\multiput(3.5,4)(0.25,0){4}{\line(1,0){0.15}}
			\multiput(8.5,1)(0.25,0){4}{\line(1,0){0.15}}
			\multiput(8.5,4)(0.25,0){4}{\line(1,0){0.15}}
			}%
			
			{\dottedLines
			\multiput(4.5,1)(0.1,0){10}{\line(1,0){0.03}}
			\multiput(9.5,1)(0.1,0){10}{\line(1,0){0.03}}
			\multiput(4.5,4)(0.1,0){10}{\line(1,0){0.03}}
			\multiput(9.5,4)(0.1,0){10}{\line(1,0){0.03}}
			}%
			
			\multiput(2,4)(5,0){2}{\circle{1.3}}
			\multiput(2,1)(5,0){2}{\circle{1.3}}
			\multiput(4.5,2.5)(5,0){2}{\circle{1.3}}
			
			\multiput(0.5,4)(1,0){11}{\circle*{0.1}}
			\multiput(0.5,1)(1,0){11}{\circle*{0.1}}
			\multiput(4.5,2)(0,1){2}{\circle*{0.1}}
			\multiput(9.5,2)(0,1){2}{\circle*{0.1}}
			
				\put(1.77,3.85){$\G_1$}
				\put(6.77,3.85){$\G_4$}
				\put(4.27,2.35){$\G_3$}
				\put(9.27,2.35){$\G_6$}
				\put(1.77,0.85){$\G_2$}
				\put(6.77,0.85){$\G_5$}
			
			\put(-0.5,5.6){$x=$}
			\put(0.5,5.5){\vector(0,-1){1}}
			\put(0.375,5.6){$0$}
			\put(3.5,5.5){\vector(0,-1){1}}
			\put(3.4,5.6){$1$}
			\put(4.5,5.5){\vector(0,-1){1}}
			\put(4.4,5.6){$2$}
			\put(5.5,5.5){\vector(0,-1){1}}
			\put(5.375,5.6){$3$}
			\put(8.5,5.5){\vector(0,-1){1}}
			\put(8.35,5.6){$4$}
			\put(9.5,5.5){\vector(0,-1){1}}
			\put(9.375,5.6){$5$}
			\put(10.5,5.5){\vector(0,-1){1}}
			\put(10.4,5.6){$6$}
		\end{picture}
\caption{\label{fig:schemeLayout}
(Color online) Implementation of single-qubit
operations.  The solid green, dashed red, and dotted blue lines form three
distinct disconnected graphs, $\Gsolid$, $\Gdashed$, and $\Gdotted$,
respectively;
these graphs are switched on and off via global control in an
algorithm-independent sequence, namely: $\solid,\dashed,\solid,\dotted$,
repeat.  The graphs $\G_{j}$, $j\in\Z$, are determined by the desired
algorithm, and are chosen from a specified set.  The result is that a
walker initially at $x=0$ will take discrete steps to successive values of
$x$, being transformed in the process.}
\end{figure}

Fig.~\ref{fig:schemeLayout} shows all of these elements combined to form a
single-qubit gate via hybrid discrete-continuous quantum walk.  The graphs
$\G_j$ are to be chosen from a universal set of graphs that we determine in
Section \ref{sec:blocks}.  They are such that when the transport rails attached
to them are turned on while the walker is at a node with position $x=3j$,
$j\in\Z$, it undergoes PST to the node at $3j+1$ and is transformed as desired
in the process.  The choice of the $\G_j$ is algorithm dependent, but once made
the graphs remain in place unchanged throughout the following protocol, which
requires a level of global control only to switch among three sets of transport
segments.

A walker is initialized on the left-most vertex of the $\ket{0}$ rail and the
solid transport rail segments, labeled by \Gsolid, are enabled.
(Note that at any given time only one
transport graph -- \Gsolid, \Gdashed, or \Gdotted\ -- is enabled, so when one
is stated to be on the others are implied to be off.)  After a time $t_h$, which
must be the same for all $\G_j$ to which horizontal rail segments attach, the
walker has unit probability to be at $x=1$.  The next step is taken with
\Gdashed\ enabled for a time $t_m=\pi/2$,  moving the walker to $x=2$.  The 
graphs $\G_j$ for the vertical rail segments are such that after a time $t_v$, 
with \Gsolid\ enabled again, either the walker remains unchanged or is 
transformed into a superposition of the $\ket{0}$ and $\ket{1}$ rails at $x=2$.  
In either case, \Gdotted\ is the next to be turned on for a time $t_m$,
moving the walker to $x=3$, possibly spread over two rails.  This 
sequence now repeats:
\Gsolid\ for $t_h$, \Gdashed\ for $t_m$, \Gsolid\ for $t_v$,
\Gdotted\ for $t_m$.
Each iteration moves the walker three $x$ positions to the right in a time of
$t_h+t_v+\pi$, enacting operations upon it along the way.  
After traversing the whole graph, involving some number of iterations, the state 
of the walker at the output on the right will be the desired arbitrary
single-qubit state $\alpha\ket{0}+\beta\ket{1}$, with $|\alpha|^2+|\beta|^2=1$.
The next step is to expand this scheme from single-qubit operations to universal
quantum computation, which will follow from universal single-qubit computation 
plus a two-qubit entangling gate.

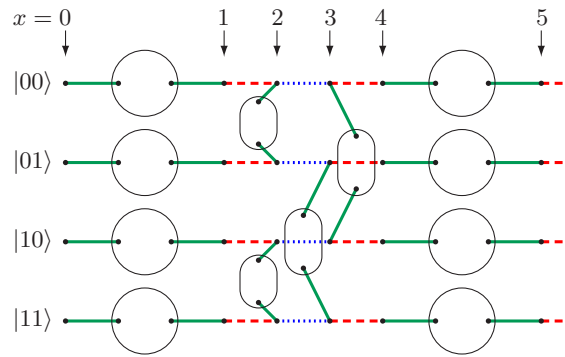
\begin{figure}
	\setlength{\unitlength}{1pt}
	\begin{picture}(220,133)(-20,-13)
		\put(-20,88){$\ket{00}$}
		\put(-20,58){$\ket{01}$}
		\put(-20,28){$\ket{10}$}
		\put(-20,-2){$\ket{11}$}
		
		{\dashedLines
			\multiput(60,0)(40,0){2}{
				\multiput(0,0)(0,30){4}{
					\multiput(0,0)(5,0){4}{\line(1,0){3}}
				}
			}
			\multiput(180,0)(0,30){4}{
				\multiput(0,0)(5,0){2}{\line(1,0){3}}
			}
		}%
		{\dottedLines
			\multiput(80,0)(0,30){4}{
				\multiput(0,0)(2,0){10}{\line(1,0){0.7}}
			}
		}%
		{\solidLines
			\multiput(0,0)(0,30){4}{\line(1,0){20}}
			\multiput(40,0)(0,30){4}{\line(1,0){20}}
			\multiput(120,0)(0,30){4}{\line(1,0){20}}
			\multiput(160,0)(0,30){4}{\line(1,0){20}}
			\multiput(80,0)(0,60){2}{\line(-1,1){7}}
			\multiput(73,23)(0,60){2}{\line(1,1){7}}
			\put(100,0){\line(-1,2){10}}	
			\put(100,30){\line(1,2){10}}
			\put(100,60){\line(-1,-2){10}}
			\put(100,90){\line(1,-2){10}}
		}%
		
		\multiput(0,0)(20,0){10}{\circle*{2}}
		\multiput(0,30)(20,0){10}{\circle*{2}}
		\multiput(0,60)(20,0){10}{\circle*{2}}
		\multiput(0,90)(20,0){10}{\circle*{2}}
		
		\multiput(30,0)(0,30){4}{\circle{25}}
		\multiput(150,0)(0,30){4}{\circle{25}}
		\multiput(73,15)(0,60){2}{\oval(14,20)}
		\multiput(73,7)(0,16){2}{\circle*{2}}
		\multiput(73,67)(0,16){2}{\circle*{2}}
		\multiput(90,30)(20,30){2}{\oval(14,25)}
		\put(90,20){\circle*{2}}
		\put(110,50){\circle*{2}}
		\put(90,40){\circle*{2}}
		\put(110,70){\circle*{2}}
		\put(-20,112){$x=0$}
		\put(0,110){\vector(0,-1){10}}
		\multiput(60,110)(20,0){4}{\vector(0,-1){10}}
		\put(57.5,112){$1$}
		\put(78,112){$2$}
		\put(98,112){$3$}
		\put(117,112){$4$}
		\put(180,110){\vector(0,-1){10}}
		\put(178,112){$5$}
	\end{picture}
	\caption{\label{fig:twoQubits}(Color online)
		Hybrid scheme for a two qubit computation.  The required sequence
		for the three transport rail types is:
		$\solid,\dashed,\solid,\dotted,\solid,\dashed$, repeat.
		Each additional qubit adds another set of either 
		\dotted,\solid\ or \dashed,\solid\ before the repeat,
		such that the
		\Gdashed\ and \Gdotted\ graphs
		alternate on the horizontal rail.}
\end{figure}

To extend our scheme to two qubits, we require four rails.  The horizontal 
portions of the protocol remain the same, however we now require vertical 
connections between additional pairs of rails at each step.  The graph that 
allows this can be seen in Fig.~\ref{fig:twoQubits}.  There is a new sequence 
for switching on and off the transport rail sets, but there are still only three 
distinct segment types required.
For $N$ qubits the number of rails required is $2^N$ and the number of
inter-rail connections at a single $x$ value (e.g.\ $x=2$ or $3$ in 
Fig.~\ref{fig:twoQubits}) is $2^{N-1}$. Note that this affects only the width 
of the graph: the number of distinct $x$ values at which these inter-rail 
connections are required at each stage in the sequence is only $N$. 
This is because to perform a single-qubit rotation on the $n$th 
qubit out of $N$, the rail corresponding to $\ket{b_1\cdots 0_n\cdots b_N}$
must be connected to $\ket{b_1\cdots 1_n\cdots b_N}$ for each of the $2^{N-1}$
arrangements of the $b_i\neq b_n$, but these connections are simultaneous.
Since single-qubit rotations are required for each qubit, we require $N$ such
sets of connections at each step, so the depth of the graph and therefore the
time taken to traverse it is linear in the number of qubits.  The number of
these inter-rail connections compares directly with the requirements of
previous schemes~\cite{Childs2009-PRL102-180501,Lovett2010-PRA81-42330}.

Given a set of graphs $\{\G_j\}$ that provides a universal gate set,
which we describe in the following section, we are now in a position to
describe the simulation of a quantum circuit on $N$ qubits,
with a depth of $D$.  We define a
layer of the computational graph to be one of three things:
(1) a horizontal widget on each rail, mediated by \Gsolid,
(2) the subgraph of \Gdashed\ or \Gdotted\ joining $x$ and $x+1$ for some $x$,
or (3) the subgraph of \Gsolid\ at a single $x$ value, providing the vertical
connections that allow a basis-changing operation on a single qubit.
We further define a horizontal sequence $\seq_h$ as the enabling of \Gsolid\ 
for a duration $t_h$ followed by \Gdashed\ for $t_m$, and a vertical sequence
$\seq_v^{(j)}$, for $j\in\{\dashed,\dotted\}$, as \Gsolid\ enabled for
$t_v$ followed by $\GdashORdot{j}$ for $t_m$.  The protocol proceeds as
follows.

Algorithm-specific graphs are inserted into the generic structure of the
transport graphs $\Gsolid$, $\Gdashed$, and $\Gdotted$.  With these graphs
disabled, a quantum walker is initialized on the first node of the
$\ket{00\cdots0}$ rail.  The transport graphs are then cycled on and off
according to
\beq{eq:graphSequence}
	\seq_h,\underbrace{
		\seq_v^{(\dotted)},\seq_v^{(\dashed)},\ldots,\seq_v^{(\dashed)}
	}_{
		N\text{ sequences}
	},
\eeq
with the vertical sequences alternating between \dashed\ and \dotted.
This constitutes the first round of the protocol, and it finishes with the
walker at $x=2(N+1)$ after a time of $t_h+t_m+N(t_v+t_m)$. 
The set of sequences in Eq.~\eqref{eq:graphSequence} is executed a total
of $D$ times, after which the walker is in the superposition of output nodes 
corresponding to the result of the action of the circuit unitary on the input
state.  We can therefore define the `graph depth' by $D_\G=2D(N+1)$,
corresponding to the total number of operations required to simulate the
circuit of depth $D$.  Since the graph depth is polynomial in the
number of qubits, its dependence on $N$ results in at worst a logarithmic
correction factor to any quantum algorithm offering a polynomial speed-up
over the classical case.  Algorithms offering exponential speed-ups continue
to do so in this model.

If additional runs of the algorithm are required, for example to build up
statistics of the output state, they can be run almost in parallel.  
Once the first walker has reached the input node to the second round at 
position $x=N+2$, a second walker can be started at the input node of the
first round, $x=0$.  With no additional cost, the same sequence of transport
segments will then move both walkers through the computation simultaneously,
neither affected by the other's presence.
When the first walker reaches the set of final output nodes, it will remain
there at the final $x$ position while the last set of transport rails that
it traversed is off.
During this time it can be measured and ejected from the system before those
rails cycle on again.  This prevents the first walker from moving backward
into the graph toward the second one.
The whole process can of course be
repeated for further additional walkers.

The distance between walkers can in fact be made constant if the
widgets are chosen such that $t_h=t_v$ and the number of vertical sequences
required in Eq.~\eqref{eq:graphSequence} is odd, say $N=2k+1$, $k\in\Z$
\cite{ourFootnote}.
In this case, $\seq_h=\seq_v^{(\dashed)}$, and the result of
Eq.~\eqref{eq:graphSequence} for one round is simply $k+1$ copies of
the sequence $\seq_v^{(\dashed)}\seq_v^{(\dotted)}$.  Therefore a
second walker can be initialized at $x=0$ after the first of these,
i.e.\ when the first walker has reached $x=4$, independent of $N$.
For example, with $N=3$ qubits the sequence for two walkers is
\beq{eq:multipleWalkerSeqs}
	\begin{array}{rc}
		\text{\footnotesize Walker 1} & \rightarrow \\
		\text{\footnotesize Graphs:} &\vphantom{\displaystyle{\frac{1}{2}}}\\
		\text{\footnotesize Walker 2} & \rightarrow
	\end{array}
	\ 
	\overbrace{
		\solid,\dashed,\solid,\dotted,
		\makebox[0pt][l]{$\displaystyle{
			\underbrace{
				\phantom{\solid,\dashed,\solid,\dotted,\solid,\dashed,\solid,\dotted}
			}_{
				\text{Round 1}
			}
		}$}
		\solid,\dashed,\solid,\dotted}^{\text{Round 1}}
		,\solid,\dashed,\solid,\dotted,
		\makebox[-5pt]{$\displaystyle{
			\overbrace{
				\phantom{\solid,\dashed,\solid,\dotted,\solid,\dashed,\solid,\dotted}
			}^{
				\text{Round 2}
			}
		}$}\cdots\hphantom{,\solid,\dotted\ }.
\eeq
Our universal gate set, described in the next section, is of this form
with $t_h=t_v$.

\section{Widgets\label{sec:blocks}}
The fundamental elements of this scheme are the graphs $\G_j$ that are 
connected to the rails.  In this section we describe three widgets,
each comprised of a graph attached to two transport rail segments,
that together yield a universal gate set for single-qubit operations. When 
combined, these yield the identity gate, a $Z$-rotation, and an $X$-rotation
on the encoded qubit. 

In principle, the identity gate is already built into the motion along each
rail: the state of the walker after each step is simply multiplied by a factor
of $-i$. The $N$-qubit state being represented is then unaffected by the
motion along each step, besides an unimportant overall phase.
That said, all the widgets shown in Fig.~\ref{fig:schemeLayout} are graphs 
${\mathcal G}_j$ {\it combined} with two edges, connecting to the
input and output vertices of the ${\mathcal G}_j$, respectively. The smallest
identity graph possible is therefore a single vertex, which corresponds to a 
three-site unweighted linear widget. The same widget can be obtained by dividing
the edge weights of the two-segment ($M=2$) quantum wire (cf.\ 
Fig.~\ref{fig:PSTline}) by a factor of $\sqrt{2}$. Thus the simplest identity 
gate requires a PST time $t=\sqrt{2}\left(\pi/2\right)=\pi/\sqrt{2}$, and
multiplies the state of the walker by a factor of $\left(-i\right)^2=-1$. The 
same procedure can be applied to quantum wires of arbitrary length: 
dividing the edge weights by $\sqrt{M}$ yields an identity 
widget with unit weights on the first and last edges, in a PST time
$t=\sqrt{M}\left(\pi/2\right)$ and with an overall factor 
of $\left(-i\right)^M=e^{-i M\pi/2}$ for each rail.
Note though that two wires of different length cannot be combined to create
a phase gate, since they require different times to exhibit PST.

\begin{figure}
	\setlength{\unitlength}{0.045\textwidth}
	\begin{minipage}{0.45\textwidth}
		\subfloat[]{
			\label{fig:basicBlocksP}
			\begin{picture}(3,2)
				\multiput(0.5,0.6)(2,0){2}{\circle{0.15}}
				\multiput(0.5,1.6)(2,0){2}{\circle*{0.15}}
				\put(1.3,0.3){$\mu_2$}
				\put(0,1){$\mu_1$}
				\put(2.6,1){$\mu_1$}
				\put(1.3,1.8){$\mu_3$}
				{\thicklines
					\multiput(0.575,0.6)(0,1){2}{\line(1,0){1.85}}
					\multiput(0.5,0.675)(2,0){2}{\line(0,1){0.85}}
				}
			\end{picture}
		}\hfill
		\subfloat[]{
			\label{fig:basicBlocksI}
			\begin{picture}(2.4,1)
				\multiput(0.2,0.6)(2,0){2}{\circle{0.15}}
				\put(1.2,0.6){\circle*{0.15}}
				\put(0.5,0.8){$\mu_I$}
				\put(1.5,0.8){$\mu_I$}
				{\thicklines
					\put(0.275,0.6){\line(1,0){1.85}}
				}
			\end{picture}
		}\hfill
		\subfloat[]{
			\label{fig:basicBlocksR}
			\begin{picture}(2,2)
				\multiput(1,0.6)(0,1){2}{\circle{0.15}}
				\put(1.15,1){$\mu_R$}
				{\thicklines
					\put(1,0.675){\line(0,1){0.85}}
				}
			\end{picture}
		}
	\end{minipage}
\caption{
	Basic building blocks of the hybrid scheme.
	Open circles represent the nodes shared with the transport graph
	\Gsolid.  The graphs are
	\subref{fig:basicBlocksP} the phase graph $\G_P$,
	\subref{fig:basicBlocksI} the identity graph $\G_I$,
	and \subref{fig:basicBlocksR} the rotation graph $\G_R$.
	Their edge weights are 
  $\mu_I=\sqrt{3/2}$,
  $\mu_1=5\sqrt{3}/8$,
  $\mu_2=15/8$,
  $\mu_3=21/8$, and
  $\mu_R=2\sqrt{3}$.
 }
\end{figure}
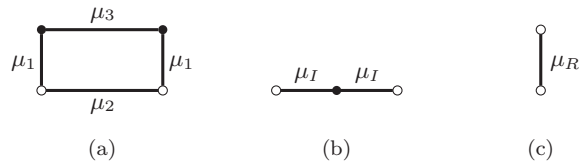

The first non-trivial gate is a $Z$-rotation, 
$R_Z(\theta)\propto\ket{0}\bra{0}+e^{i\theta}\ket{1}\bra{1}$. After a single
step of the discontinuous quantum walk, the state of the walker on the second 
rail (encoding the $\ket{1}$ component of the computational qubit) must 
accumulate a phase different by $\theta$ 
from that accumulated on the first rail. In practise, this means that the state
on the second rail must obtain a phase $\theta\neq -M\pi/2$ in a time
$t=\sqrt{M}\left(\pi/2\right)$, relative to the identity gate acting on the
first rail.

Candidate graphs on up to four vertices were considered, and within this 
restricted search space no widgets satisfied the above criteria for transit 
times $t=\sqrt{2}\pi/2$ or $\sqrt{3}\pi/2$. The first successful widget found
has transit time $t=\pi$. This is based on the graph $\G_P$ that is a weighted 
square, as shown in Fig.~\ref{fig:basicBlocksP}.  We number the vertices
clockwise around the graph, from $\ket{1}$ in the bottom left to $\ket{4}$
in the bottom right; the corresponding widget has two additional vertices,
$\ket{\vl}$ and $\ket{\vr}$, attached to $\G_P$ by transport rails on
the left and right, respectively.  The resulting widget
Hamiltonian (adjacency matrix) is
\begin{align}
	H_P
	=&\,
	\ketbra{\vl}{1}
	+\mu_1\left(\ketbra{1}{2}+\ketbra{3}{4}\right)
	+\mu_2\ketbra{1}{4} \nonumber \\
	&\quad
	+\mu_3\ketbra{2}{3}
	+\ketbra{4}{\vr}
	+\Hc\label{eq:HamiltonianP}
\end{align}
With the edge weightings
$\mu_1\equiv w_{12}=w_{34}=5\sqrt{3}/8$, $\mu_2\equiv w_{14}=15/8$, and
$\mu_3\equiv w_{23}=21/8$, the state of a walker initially on the left-hand
node $\ket{\vl}$ is transformed in a time $\pi$ as
\beq{eq:GPPST}
	\e^{-i H_P \pi}\ket{\vl}=i\ket{\vr}.
\eeq
The time for the horizontal segments of the computational graph is then taken
to be $t_h=\pi$.

To perform a $Z$-rotation based on the graph ${\mathcal G}_P$, one requires an 
identity gate taking a time $t_h=\pi=\sqrt{4}\pi/2$ on the first rail. 
Evidently this corresponds to a four-segment ($M=4$) quantum-wire widget, 
with the weights of the first and last segments rescaled to unity. The phase
acquired by the walker during traversal is $(-i)^4=1$. This 
results from the graph $\G_I$, shown in Fig.~\ref{fig:basicBlocksI}, connected 
to a rail on either end.  Since there are four line segments in total,
the weighting of the second 
and third segments should be $\mu_I=\sqrt{3/2}$. The Hamiltonian 
corresponding to this widget is
\beq{eq:HamiltonianI}
	H_I
	=
	\ketbra{\vl}{1}
		+\mu_I\sum_{j=1}^2\ketbra{j}{j+1}
		+\ketbra{3}{\vr}
	+\Hc,
\eeq
where $\{\ket{1},\ket{2},\ket{3}\}$ is the set of nodes in $\G_I$,
labeled from left to right.

The final graph we require does not actually exhibit PST.  Instead, 
with \Gsolid\ enabled the rotation graph $\G_R$ results in a widget
connecting a vertex $\ket{\vt}$ on the top to $\ket{\vb}$ on
the bottom of a pair of rails.  In a time $t_v=\pi$ the
effect of this widget on a walker 
starting at either $\ket{\vt}$ or $\ket{\vb}$ is to split its probability
density  between these two states, leaving no probability inside the graph
itself.  $\G_R$ consists of a single weighted line segment, and can be seen
in Fig.~\ref{fig:basicBlocksR}.  Its widget Hamiltonian is
\beq{eq:HamiltonianR}
	H_R
	=
	\ketbra{\vt}{1} + \mu_R\ketbra{1}{2} + \ketbra{2}{\vb} + \Hc,
\eeq
and with weighting $\mu_R=2\sqrt{3}$, the action on walkers initially on
either the top or bottom rail is given by
\begin{align}
	\e^{-i H_R t_v}\ket{\vt}
		&= \cos(\sqrt{3}\pi)\ket{\vt} - i\sin(\sqrt{3}\pi)\ket{\vb}, \nonumber\\
	\e^{-i H_R t_v}\ket{\vb}
		&= -i\sin(\sqrt{3}\pi)\ket{\vt} + \cos(\sqrt{3}\pi)\ket{\vb}.
		\label{eq:basisChange}
\end{align}
Note that if the graph $\G_R$ is not present, which is equivalent to
setting the weight $\mu_R=0$, then this same widget acts as an identity
operation in the same time $t_v$.  In this case, whether it starts at
$\ket{\vt}$ or $\ket{\vb}$ the walker sees only a single line segment.
It walks the line in time $t_v/2$, acquiring a phase of $-i$.
Therefore after an elapsed time of $t_v$ the walker has made a round
trip and returned to its initial position with an accumulated phase of
$(-i)^2=-1$; its state is unchanged, up to a global phase.

\section{Universal computation\label{sec:universal}}
We now show how the graphs $\G_I$, $\G_P$, and $\G_R$ can be combined to
construct a universal set of gates for single-qubit operations, and then
add a controlled-$Z$ gate to provide universal quantum computation.

\begin{figure}
\setlength{\unitlength}{1cm}
		\begin{picture}(5.6,2.8)(0,0.2)
			
			\put(0,2.5){$\ket{0}$}
			{\solidLines
				\put(0.5,2.6){\line(1,0){1}}
				\put(3.5,2.6){\line(1,0){1}}
				\put(0.5,0.6){\line(1,0){1}}
				\put(3.5,0.6){\line(1,0){1}}
			}%
			
			{\thicklines
				\put(1.5,2.6){\line(1,0){1}}
				\put(2.5,2.6){\line(1,0){1}}
			}
			
			\put(0.5,2.6){\circle*{0.1}}
			\put(1.5,2.6){\circle*{0.1}}
			\put(2.5,2.6){\circle*{0.1}}
			\put(3.5,2.6){\circle*{0.1}}
			\put(4.5,2.6){\circle*{0.1}}
			\put(4.75,2.5){$\ket{0}$}
			\put(1.75,2.8){$\mu_{I}$}
			\put(2.75,2.8){$\mu_{I}$}
			
			\put(0,0.5){$\ket{1}$}
			{\thicklines
				\put(1.5,0.6){\line(1,0){2}}
				\put(1.5,0.6){\line(0,1){1}}
				\put(1.5,1.6){\line(1,0){2}}
				\put(3.5,1.6){\line(0,-1){1}}
			}
			\put(0.5,0.6){\circle*{0.1}}
			\put(1.5,0.6){\circle*{0.1}}
			\put(1.5,1.6){\circle*{0.1}}
			\put(3.5,1.6){\circle*{0.1}}
			\put(3.5,0.6){\circle*{0.1}}
			\put(4.5,0.6){\circle*{0.1}}
			\put(4.75,0.5){$\e^{i \frac{\pi}{2}}\ket{1}$}
			\put(1.1,1){$\mu_{1}$}
			\put(3.6,1){$\mu_{1}$}
			\put(2.25,0.35){$\mu_{2}$}
			\put(2.25,1.75){$\mu_{3}$}
			
		\end{picture}
\caption{\label{fig:rootZ}(Color online) Single-qubit $\sqrt{Z}$ gate.}
\end{figure}
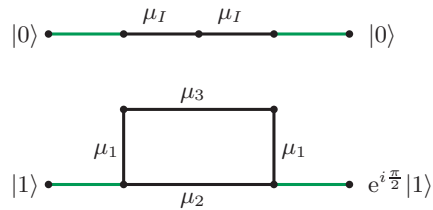

The $\sqrt{Z}$ phase gate is straightforward to construct. Consider only the 
first stage of the graph in Fig.~\ref{fig:schemeLayout}, comprised of the 
graphs $\G_1$ and $\G_2$ along with the connector segments that attach them to 
the $x=0$ and $x=1$ nodes.
As shown in Fig.~\ref{fig:rootZ}, we replace $\G_1$ with the identity graph 
$\G_I$, and substitute the phase graph $\G_P$ for $\G_2$. A walker starting 
in an arbitrary superposition of computational basis states on the left-most
vertices of the gate,
$\ket{\psi(t=0)}=\alpha\ket{0}+\beta\ket{1}$, will after a time $t_h$ be in 
the state $\ket{\psi(t=t_h)}=\alpha\ket{0}+\beta\e^{i\pi/2}\ket{1}=
\sqrt{Z}\ket{\psi(t=0)}$ on the right-most vertices.
Note that placing $\G_P$ on the $\ket{0}$ rail and $\G_I$ on the $\ket{1}$ rail
changes the resulting gate into $i\sqrt{Z}^\dagger$, and of course putting 
$\G_I$ on both rails results in the identity gate, $I$.

After enacting $\sqrt{Z}$, $i\sqrt{Z}^\dagger$, or $I$, the walker has moved 
from $x=0$ to $x=1$.  It is then transferred to $x=2$ via PST across a single 
line segment. At this point a basis change can be effected, if desired, by
using the rotation graph $\G_R$ with the states $\vin$ and $\vout$ of 
Fig.~\ref{fig:basicBlocksR} identified with the logical rail states $\ket{0}$
and $\ket{1}$, respectively.  Using Eq.~\eqref{eq:basisChange} one readily 
obtains
\beq{eq:basisChangingGate}
	\ket{\psi(t=t_v)}=R_X(2\sqrt{3}\pi)\ket{\psi(t=0)}.
\eeq

The $\sqrt{Z}$ and $R_X(2\sqrt{3}\pi)$ gates constitute a universal set for 
single-qubit operations. Conjugating the latter by the former gives
\beq{eq:genRY}
	\sqrt{Z}R_X(2\sqrt{3}\pi)\sqrt{Z}^\dagger
	=
	R_Y(2\sqrt{3}\pi),
\eeq
i.e.\ a $Y$ rotation through an angle $2\sqrt{3}\pi$.
As with Euler-angle rotations in three-dimensional Cartesian space
and due to the correspondence between $SO(3)$ and $SU(2)$,
these rotations of the Bloch sphere by irrational multiples of $\pi$ about 
non-parallel axes allow an arbitrary rotation to be performed, and therefore
provide a universal set of single-qubit gates.

All that remains is to construct a two-qubit entangling gate. The 
implementation of $\sqrt{Z}$, by the application of a phase to a single computational 
basis state, suggests a straightforward method for implementing the entangling 
controlled-$\sqrt{Z}$ gate, $\sqrt{CZ}$.
With rails for two qubits we simply apply $\sqrt{Z}$ to the $\ket{11}$
rail while applying identity operations to the 
$\ket{00}$, $\ket{01}$, and $\ket{10}$ rails, thus applying a phase of 
$\e^{i\pi/2}$ to $\ket{11}$ relative to the other three computational 
basis states.  Repeating this obviously results in a full $CZ$ operation.
In combination with the 
universal set of single-qubit operations already described, the ability to 
implement a $CZ$ gate makes this scheme universal for quantum 
computation.

\section{Conclusions\label{sec:summary}}
By combining components of perfect state transfer and quantum walks we have
developed a hybrid scheme for performing universal quantum computation 
via a walker taking discrete steps of continuous evolution, a `discontinuous
quantum walk.' The computational 
model is based on one rail per computational basis state, as developed for 
prior schemes to provide universal quantum computation in the distinct cases 
of continuous~\cite{Childs2009-PRL102-180501} and 
discrete~\cite{Lovett2010-PRA81-42330} walks. As in the discrete case,
we have eliminated the need for the excess tails used in the continuous case to
support well-defined momentum states, and do not require the momentum filter 
that prevents most of the walker from participating in the computation.  By 
making use of perfect state transfer, we ensure that the walker completes the 
quantum computation with certainty.  Unlike the discrete case, the evolution of 
our quantum walker is manifestly physical under a specific Hamiltonian, 
and we do not require site-dependent coins of multiple dimensions or indeed
any coin at all.  The cost associated with these improvements is an additional 
amount of global control, that is analogous to the coin and shift operations 
employed by discrete-time quantum walks with site-independent coins. The 
required control is algorithm independent, conforming to a well-defined, 
pre-programmed sequence.

The widgets described in Sec.~\ref{sec:blocks} are universal for quantum
computation, so they provide a proof-of-principle scheme for the implementation 
of arbitrary quantum algorithms. That said, they are neither unique nor or 
they likely to be a preferred set for particular applications. Alternative
choices of single-rail and double-rail graphs (generating single-qubit gates)
might generate particular desired gates (such as the Hadamard or $\pi/8$-gate)
more readily. Multi-qubit gates (such as the three-qubit Toffoli gate) could
be found by graphs linking multiple rails. A desired unitary for $n$ qubits 
would conceivably have a more efficient decomposition in terms of a larger 
widget set.  This decomposition would be in the same spirit as the model
employed in Ref.~\cite{Cleve-STOC} to examine the relationship between
discrete and continuous quantum query algorithms, but would require
neither so-called fractional queries nor Trotter-Suzuki type approximations.

Regardless of the choice of widgets one can recast their
Hamiltonians in terms of spin networks, in the spirit of
Ref.~\cite{Christandl2005-PRA71-032312}, perhaps providing a closer link
to potential experimental implementations.  This is possible because
a quantum walker on a $k$-vertex
graph can be mapped onto the single-excitation subspace of
a system of $k$ spin-$\frac{1}{2}$ particles under the $XY$ model.
The spin-preserving Hamiltonian of this model is of the form
$H\sim\frac{1}{2}\sum_j\left(X_jX_{j+1}+Y_jY_{j+1}\right)$.
In this context the particles
themselves remain stationary and take the place of the nodes, while the
exchange interaction provides edges along which the
excitation propagates.
The correspondence between the $XY$ and quantum-walk models can be
seen directly in the behavior of two interacting spins: an excitation on
the left spin evolves to an excitation on the right one.
This is nothing but a Pauli $X$ operation, as effected on a quantum
walker under the influence of
the hopping Hamiltonian on the two-vertex connected graph.

More generally, the discontinuous quantum walk provides a framework for 
universal control of a quantum system. Though the universality of quantum
computation is presented above in analogy to the circuit model, with rails 
corresponding to computational basis states, this is not in fact essential. In 
principle, the edges between subgraphs of any particular graph can be turned 
on and off in a prescribed manner, in the process effecting some desired 
operation on the quantum walker. The total number of vertices would still 
presumably scale exponentially in the number of simulated qubits, but the 
representation of the graph for some quantum algorithms could be much more 
efficient than that proposed above.  We hope that the flexibility of the 
discontinuous quantum walk will lend itself naturally to the development of new
efficient quantum algorithms.

\section*{Acknowledgments}
We are grateful to Andrew Childs, Barry Sanders, and Peter
H\o yer for fruitful discussions during the preparation of this work.
This work was supported by Natural Sciences and Engineering Research Council of
Canada (NSERC) and the Alberta Informatics Circle of Research Excellence 
(iCORE).


\begin{thebibliography}{20}
\bibitem{ourFootnote}{As we have
described the protocol, an odd number of vertical sequences corresponds
to an odd number of qubits.  If an even number of qubits is desired
along with a constant distance between walkers, an additional vertical
identity sequence could simply be added.}
\end{thebibliography}
\end{document}